# DETERMINATION OF VACANCIES ALLOCATION IN A MONOCRYSTAL OF A P-DICHLOROBENZENE USING A METHOD OF THE RAMAN SPECTROSCOPY

## M.A. Korshunov[*]

Kirensky Institute of Physics, Siberian Division, Russian Academy of Sciences, Krasnoyarsk, 660036 Russia

**Abstract**. The method of Raman scattering is used to investigate the cross edge of a p-dichlorobenzene monocrystal, which was grown up by a Bridgmen method. Comparison of intensities of additional lines of a spectrum of the lattice oscillations of certain parts of an edge of a monocrystal with density of these fields has allowed to relate some lines to presence of vacancies in structure.

Dynamics of a lattice of real crystals is influenced by presence of impurities, in particular vacancies. It discovers the reflexion in spectrums of a Raman effect of light of small frequencies. Therefore the method Raman of spectroscopy is rather convenient tool of diagnostics and optical monitoring. Properties of monocrystals depend, as vacancies are proportioned on monocrystal volume. For examination of it the monocrystal of a p-dichlorobenzene has been grown up by a method of the Bridgmen. Further definition of concentration of builders on diameter of a monocrystal was spent. The grown up monocrystal on length was slited on tablets by thickness ~0.3 see Then some parallelepipeds were cut out from these tablets for examination of radial - allocation of an impurity from centre to edge with a size of an edge 0 the Monocrystallinity of samples see was checked by means of a polarizable - microscope. Spectrums of explored - samples have been gained Raman. For each sample the density has been measured. To define a vacancy concentration it is possible comparing the measured density of a crystal with the density calculated on X-ray diffraction data.

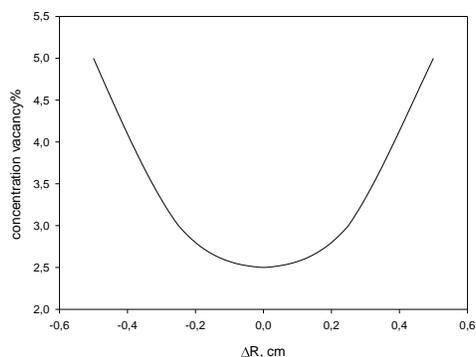

Fig. 1.

---

[*] mkor@iph.krasn.ru

In Figure 1 the vacancy concentration modification on diameter of a monocrystal of a p-dichlorobenzene is shown. As we see to edges of the sample a vacancy concentration increases.

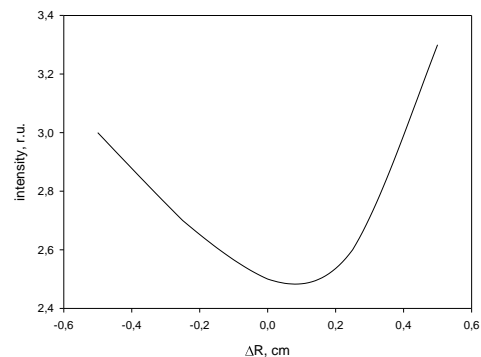

Fig. 3.

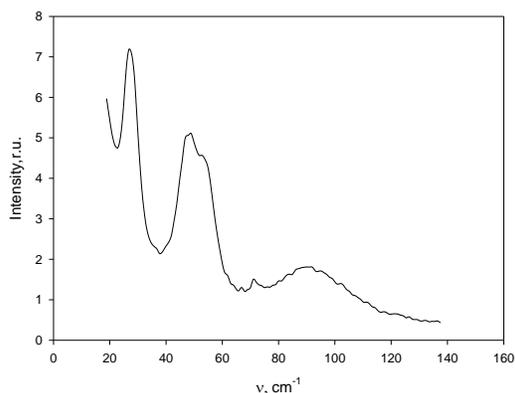

Fig. 2.

In Figure 2 the example of a spectrum of the lattice oscillations of a p-dichlorobenzene is shown. In a spectrum it is observed six intensive lines of the molecules caused by orientation oscillations. Additional lines of a small intensity in the field of 70 cm$^{-1}$ are related as show calculations of spectrums of the lattice oscillations with presence in structure of vacancies [1].

In Figure 3 the diagramme of dependence of an intensity of a line of ~70 cm$^{-1}$ on diameter of a monocrystal of a p-dichlorobenzene is shown. Thus an intensity of lines caused by orientation oscillations does not vary almost. As we see, comparing diagrammes 1 and 3, they are similar. It is possible to guess, that an intensity of this line is related to a vacancy concentration, and also occurrence of this line in the field of 70 cm$^{-1}$ is caused by presence of vacancies in a monocrystal.

Thus, it is possible to guess, that occurrence of additional lines of a small intensity in the field of 70 cm$^{-1}$ of a p-dichlorobenzene is caused by presence of vacancies. The vacancy concentration to boundaries of a cross edge of a monocrystal raises. On - visible allocation of vacancies in the sample is not uniform.